\documentclass{emulateapj}

\slugcomment{Not to appear in Nonlearned J., 45.}

\shorttitle{Dwarf galaxies in galaxy clusters}
\shortauthors{S\'anchez-Janssen et al.}

\begin{document}


\title{Properties of the Dwarf Galaxy Population in Galaxy Clusters}


\author{R. S\'anchez-Janssen\altaffilmark{1}, J. Alfonso L. Aguerri\altaffilmark{1} and C. Mu\~noz-Tu\~n\'on\altaffilmark{1}}
\altaffiltext{1}{Instituto de Astrof\'{\i}sica de Canarias. C/ V\'{\i}a L\'actea s/n, 38200 La Laguna, Spain}


\begin{abstract}
We present the observational properties of the dwarf galaxy population ($M_{r}>M_{r}^{*}+1$) corresponding to one of the largest samples of spectroscopically confirmed galaxy cluster members reported in the literature.  We have observed that red dwarf galaxies ($u-r>2.22$) share the same cluster environment as the brightest cluster members ($M_{r}<-21$), but are not in dynamical equilibrium. We computed the dwarf-to-giant ratio (DGR) using a spectroscopically selected sample. The DGR was found to vary with clustercentric distance, essentially due to the blue dwarf population ($u-r<2.22$). The $u-r$ color of red dwarf galaxies was independent of their environment and similar to the color of red isolated dwarfs. Blue dwarf galaxies located outside $r_{200}$ show similar $u-r$ colors to those of the field population, while strong reddening was observed toward the cluster center. We also present evidence  that the fraction of red to blue dwarf galaxies in clusters is larger in the innermost cluster regions. We conclude that the present red dwarf population observed in the central regions of nearby galaxy clusters could be related to the blue dwarf population observed in clusters at high redshift. 
\end{abstract}


\keywords{galaxies: clusters: general --- galaxies: evolution --- galaxies: dwarf}



\section{Introduction}

The hierarchical galaxy formation scenario proposes that massive galaxies grow continously through the agglomeration of smaller ones (White \& Rees 1978). In this framework dwarf galaxies would then be the building blocks of the galaxy formation process. Nevertheless, their origin and evolution are still a  matter of debate. In the standard scenario, dwarf galaxies are objects formed  via the gravitational collapse of primordial density fluctuations. Once the first stars are formed, mechanisms of energy feedback into the interstellar medium are proposed in order to regulate the subsequent star formation or even to change the structure of the galaxy \citep[][]{dekel86}, a scenario supported by several observational evidences \citep[e.g.][and references therein]{derijcke05}. However, there are also several observational proofs against the primordial origin of present-day dwarf galaxies in clusters. We can mention, for example: the large scatter found in  metallicities and ages of dwarf galaxies in the Virgo and Fornax clusters \citep[][]{conselice02}; the different faint-end slope of the galaxy luminosity function (LF) in clusters and in the field (Blanton et al. 2005; Popesso et al. 2006); the variation of the dwarf-to-giant ratio (DGR) as a function of clustercentric distance observed in some nearby cluster \citep[][]{phillips98}; or the preferential location of red bright and dwarf galaxies in overdense regions of the Universe, in contrast to blue ones \citep[][]{hogg04}.

In high density environments such as galaxy clusters, there are many different physical mechanisms that can influence the evolution of cluster galaxies; e.g, harassment \citep[][]{moore96}, or ram pressure stripping \citep[][]{quilis00}. In this scenario, dwarf galaxies can be formed and destroyed, especially in the innermost regions of galaxy clusters, where strong tidal forces can produce the destruction of galaxies, especially dwarfs \citep[][]{merritt84}. In the Coma cluster, it has been observed that dwarf galaxies do not reach the very central regions of the cluster, being distributed in a shell around the cluster center \citep[][]{trujillo02}. Dwarf galaxies can also be created in galaxy clusters due to the evolution of brighter ones.  Strong tidal interactions between galaxies and with the cluster potential can result in important mass loss and transform bright galaxies into dwarfs (Mastropietro et al. 2005; Aguerri et al. 2004, 2005a).

In this study we have measured the main observables of the dwarf galaxy population in one of the largest samples of spectroscopically confirmed galaxy cluster members. The data will be presented in \S \ref{sec:sec2}; the results on the main dwarf galaxies characteristics are given in \S \ref{sec:sec3}, and in \S \ref{sec:sec4} we present the discussion and conclusions.

\section{The Data}
\label{sec:sec2}
A detailed description of the data, cluster global parameters and cluster membership is given in Aguerri et al. (2007) (hereafter Paper I). We present here a brief summary. The data comprise all nearby clusters with known redshift at $z<0.1$ from the catalogues of Abell et al. (1989), Zwicky et al. (1961), B{\"o}hringer et al.(2000), and Voges et al. (1999) that were mapped by the SDSS-DR4 \citep[][]{york00}.  We downloaded only those galaxies located within a radius of 4.5 Mpc around the centers of the galaxy clusters.\footnote{Throughout this study we have used the cosmological parameters $H_{0}=75$ km s$^{-1}$ Mpc$^{-1}$, $\Omega_{m}=0.3$ and $\Omega_{\Lambda}=0.7$.} Only those isolated clusters with more than 30 galaxies with spectroscopic data within the search radius and high spectroscopic completeness ($>70\%$) were considered. We adopted the same isolation criterion for clusters as Biviano \& Girardi (2003; see PaperI).
 
Cluster membership was decided using the Zabludoff et al. (1998)  method (ZHG) as a first approximation and the KMM algorithm \citep[][]{ashman94} for the final values (see PaperI). We computed the mean velocity ($V_{c}$) and velocity dispersion ($\sigma_{c}$) of the galaxies belonging to each cluster.  Following the same approximation as Carlberg et al. (1997), we also determined for each cluster the radius (r$_{200}$) where the inner density is 200$\rho_{c}$, $\rho_{c}$ being the critical density of the Universe.


The final catalog contains 89 isolated galaxy clusters. As mentioned before, the data were downloaded from the SDSS-DR4 database according to a metric criterion. This means that we are mapping different physical regions for each cluster. In order to avoid possible biases, we have studied the r$_{max}$/r$_{200}$ ratio for each cluster, where r$_{max}$ is the maximum distance of a galaxy from its cluster center. We have found that all the clusters in our sample reach r$_{max}$/r$_{200}$=2, and that 50$\%$ of them reach r$_{max}$/r$_{200}$=5. The final sample of galaxies consist of 6880 galaxies located within a radius $2\times$r$_{200}$, and 10865 within 5$\times$r$_{200}$. We builded an ensemble cluster by normalizing the scales  and velocities of each galaxy. Thus, the radial distance of each galaxy to the cluster center was scaled by the r$_{200}$ of the corresponding cluster, and the relative velocity of each cluster galaxy was normalized by the velocity dispersion of the cluster.

All galaxy magnitudes were corrected for Galactic absorption and $k$-corrected to the rest-frame at $z=0$. The galaxies were classified according to their $u-r$ color, as described by Strateva et al. (2001). They showed that the distribution of SDSS galaxies in color--color diagrams is strongly bimodal, with an optimal color separation of $u-r=2.22$ for early (red) and late-type (blue) galaxies.

\section{Results}
\label{sec:sec3}


Fig.~1 shows the median clustercentric distance of the galaxies as function of their absolute rest-frame $r$-band magnitude for objects residing within r $<2\times$r$_{200}$.\footnote{We have considered the dwarf population as those galaxies with absolute $r-$band magnitude fainter than $M_{r}^{*}+1$, where $M^{*}_{r}-5 log(h)=-20.04$, Blanton et al. (2005)} The 25\% and 75\% quartiles of the distributions are also overplotted. Red galaxies always have smaller median clustercentric distances  than blue ones. The median positon of galaxies brighter than $M_{r} \approx -21.0$ is smaller towards brighter magnitudes. In contrast, for the dwarf population, fainter objects are located closer to the cluster centers. This trend is the same for red and blue galaxies.

We have measured the local galaxy surface density by taking into account the 10 nearest neighbors to each galaxy. Figure~1 shows the behavior of the projected local galaxy density as a function of the rest-frame absolute $r$-band magnitude. It can be noticed that red galaxies are always located in denser environments than blue ones. Galaxies brighter than $M_{r}=-21.0$ are at denser environments as they are brighter, and dwarf galaxies are located in denser environments as they are fainter. The two previous relations reflect the well known morphological segregation for bright galaxies in clusters, i.e., early-type galaxies are located closer to the cluster center and in denser environments than late-type ones. Binggeli et al. (1987) and Lisker et al. (2006) also found that early-type dwarf galaxies in the Virgo cluster are located in dense environments together with bright E galaxies.

The correlation between the velocity dispersion and the absolute $r$-band magnitude of the galaxies is also shown in Fig.~1. There is a segregation between red and blue objects: red ones show lower velocity dispersions than blue ones. The brightest objects ($M_{r}<-21.0$) show a smaller velocity dispersion as they are brighter. Galaxies with $M_{r}>-21.0$ have a constant velocity dispersion independent of their absolute magnitude. Similar correlations were found previously with other cluster samples \citep[][]{biviano04,goto05}.

\begin{figure}[!htb]
\centering
\includegraphics[angle=0,scale=1.0]{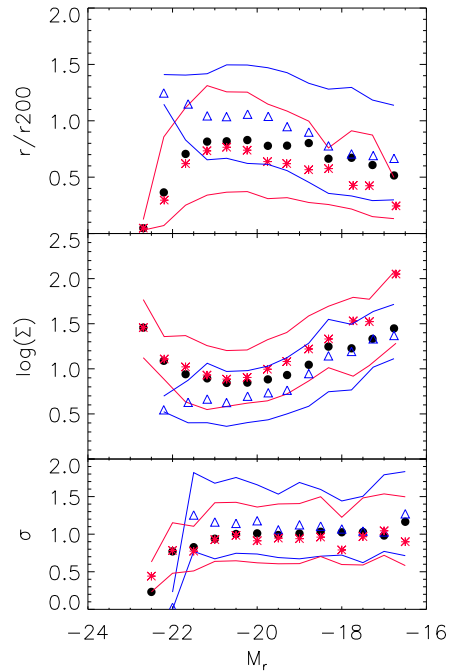}
\caption{Median position (top), local projected density (middle) and velocity dispersion (bottom) of galaxy cluster members as a function of their $r$-band absolute magnitude. In all panels, black, blue and red symbols represent the median values of these quantities for the total, blue ($u-r<2.22$) and red ($u-r>2.22$) galaxy populations, respectively. We have also overplotted the 25$\%$ and 75$\%$ quartiles for the blue and red galaxy population.}
\end{figure}

We repeated the same calculations with galaxies located at r $<5\times$r$_{200}$ in order to analyze the influence of the incompleteness in the mapping of the clusters, obtaining the same trend in all relations. Another important incompleteness is the loss of faint galaxies towards higher redshifts. Our galaxy cluster sample is complete for galaxies brighter than $M_{r}=-20.0$ (see Fig.~5 in Paper I). In order to investigate how this incompletness influences our results we have measured the median position, the projected surface galaxy density and the velocity dispersion for those galaxies located in clusters with $z<0.055$. This cluster subsample is complete for galaxies with $M_{r}<-18.5$. No changes in the trend of the relations were discovered.

The galaxy density inside $r_{200}$ in our clusters shows large variations. This large cluster richness variation could be responsible for the scatter of the measured quantities shown in Fig.~1. In order to account for this, the sample was split into two (high and low galaxy density clusters) and Fig.~1 was recomputed. Nevertheless, both samples showed  similar trends and scatter in the quantities reported in Fig.~1. This implies that the scatter shown in Fig.~1 was intrinsic and  not due to the ensemble of clusters with different richness.

We have computed the DGR as a function of clustercentric radial distance. The ratio was obtained considering as dwarf galaxies as those with $M_{r}>M_{r}^{*}+1$ and as bright ones those with $M_{r}<M_{r}^{*}$. The DGR has only been computed using  the previously mentioned complete subsample of clusters ($z<0.055$).  As can be seen in Fig.~2, the DGR increases from the innermost cluster regions towards the outermost ones. We have computed the DGR for blue and red dwarf galaxies independently. It can be seen from Fig.~2 that the DGR of red dwarf galaxies is constant, within the errors, at all radial distances. In contrast, the DGR for blue dwarfs shows a clear increase from the innermost regions towards the outermost ones. This means that the differences in DGR between the innermost regions  and the outermost ones  is basically due to blue dwarf galaxies.

The variation in the DGR with the clustercentric radial distance could be due to a variation in the density of dwarfs, giants, or both. Figure~2 shows the radial variation of the surface densities of bright and dwarf galaxies. The surface density of the bright and red dwarf galaxies show similar slopes, while the slope of the surface density of blue dwarf galaxies is larger. This means that red dwarf galaxies are more centrally concentrated than blue ones. Figure~2 also shows the ratio between red and blue dwarf galaxies as a function of radius. In the innermost regions of the cluster $r < (0.3-0.4)~r_{200}$ there are more red galaxies than blue ones. In contrast, blue galaxies are mostly located in the outermost regions ($r > r_{200}$). Previous findings \citep[][]{phillips98} have shown that dwarf galaxies avoid high density environments. We show for the first time the DGR from spectroscopic observations and conclude that it is only the blue dwarf population that avoids cluster centers. In contrast, the fraction of red dwarfs is constant (outside $r/r_{200} > 0.2$). This is in agreement with the recent result found by Hogg et al. (2004), who point out that bright and dwarf red galaxies are located in overdense regions of the Universe, while blue galaxies avoid them. As suggested by Popesso et al. (2006), there is some mechanism  transforming blue dwarf galaxies into red ones in the inner regions of clusters.

\begin{figure}[!htb]
\centering
\includegraphics[angle=0,scale=0.9]{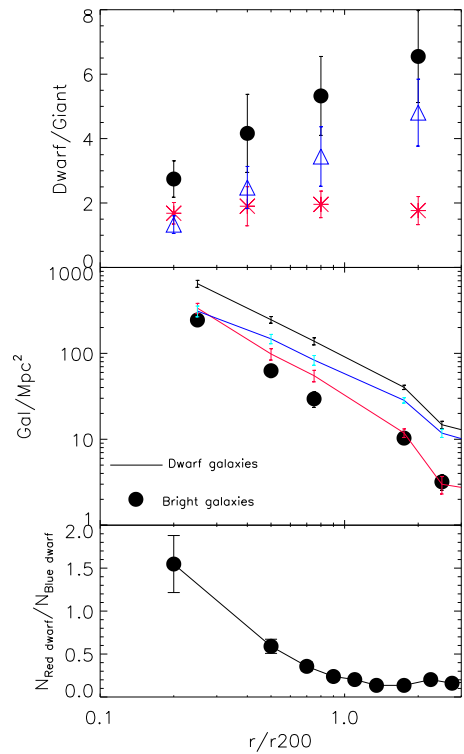}
\caption{(Top) dwarf-to-giant ratio, (middle) surface galaxy density and (bottom) red-to-blue dwarf galaxy ratio as a function of clustercentric distance. Blue and red colors as in Fig.~1.}
\end{figure}


We have also investigated the variation in the rest-frame $u-r$ color of our dwarf population as a function of the clustercentric radial distance. For comparison, we measured the mean rest-frame $u-r$ color of a sample of isolated dwarf galaxies (Allam et al. 2005). Figure~3 shows the $u-r$ color of blue and red dwarf galaxies as a function of radius. Red dwarf galaxies in clusters have a constant $u-r$ color at all radii, and similar to the mean color of isolated red dwarf galaxies. In contrast, the $u-r$ color of blue dwarf galaxies depends on the position of the galaxy in the cluster: galaxies located in the innermost regions of the cluster show redder $u-r$ colors than those located in the outermost regions. More precisely, blue dwarf galaxies located at $r < r_{200}$ show redder colors than isolated blue dwarfs, while those located at $r > r_{200}$ have similar $u-r$ colors to isolated ones (see Fig.~3).

\begin{figure}[!htb]
\centering
\includegraphics[angle=0,scale=0.5]{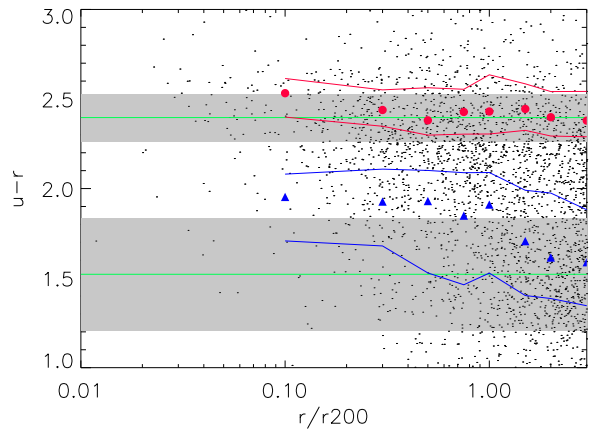}
\caption{$u-r$ color as a function of clustercentric distance of the dwarf galaxy cluster members (small points). Circles and triangles represent the mean color corresponding to the red and blue dwarf galaxy populations, respectively. Their 25$\%$ and 75$\%$ quartiles are also overplotted. The shaded areas represent the $1\sigma$ color distribution of red and blue dwarf field galaxies.}
\end{figure}

\section{Discussion and Conclusions}
\label{sec:sec4}

We grouped the galaxies in clusters into three different classes according to their absolute magnitude. The first class was formed by the brightest cluster galaxies ($M_{r}<-21.0$). The second class of galaxies represents those with $-21.0<M_{r}<M^{*}+1$, and the third class represents the faintest galaxies in our sample, those with $M_{r}>M_{r}^{*}+1$, which have been called dwarf galaxies. The results presented in the previous sections show that the brightest galaxies have a segregation with magnitude in velocity and position: they show the lowest velocity dispersion and are located closer to the cluster center in high density environments. As pointed out by \citep[][]{biviano04}, these galaxies are not in equilibrium within the cluster potential. No other segregation with magnitude in the velocity space was found for the other two galaxy classes, except that blue galaxies have a larger velocity dispersion than red ones. This trend has usually been interpreted in the past as a consequence of the different types of orbits presented by these two classes of galaxies, with blue galaxies lying on more radial orbits than red ones  \citep[][]{biviano04}.

Dwarf galaxies also show a segregation with magnitude and position. They are located closer to the cluster center as they are fainter, with red dwarf galaxies located at higher density environments than the blue ones. Therefore, the red dwarf population shares the same cluster environment as the brightest ones. Recent observations have shown that galaxy clusters at $z\approx 0.8$ are deficient in red dwarf galaxies in their innermost regions (Kodama et al. 2004; De Lucia et al. 2007). Indeed, the red DGR in the central regions of massive clusters has roughly doubled over the last 4 Gyr (Stott et al. 2007).

We can argue that the red dwarf population observed in the innermost regions of nearby clusters has a primordial origin.  Nevertheless, this argument is unlikely because cluster centers are places were tidal forces are very efficient for disrupting galaxies \citep[][]{merritt84}. Galaxies located in these inner regions have passed through the center many times during the cluster life, and would probably be disrupted. This scenario is also supported by the analysis of the color--magnitude relation and the stellar populations of cluster galaxies in Virgo and Fornax \citep[][]{conselice02,rakos04}.

Alternatively, we can think that present red dwarf galaxies were formed by transforming bright spiral galaxies due to interactions with the gravitational potential of the cluster or fast galaxy--galaxy encounters. Aguerri et al. (2005a) showed that, based on the photometric structural parameters of the dwarf galaxies in the Coma cluster, a small fraction of them could be identified with evolved harassed spiral galaxies. However, this transformation should produce important galaxy mass loss, which would be accumulated in the intracluster region forming the so-called intracluster light (ICL). If most of the present red dwarfs were formed by this mechanism, the intracluster medium should have a large amount of diffuse light. This is not observed in nearby clusters such as Virgo or Fornax where only $\approx 10\%$ of the light is located in the intracluster region (Arnaboldi et al. 2002; Aguerri et al. 2005b). Moreover, recent numerical simulations of ICL formation propose that this cluster component is mainly produced in major merger events that take place during the formation of the brightest cluster galaxies. Mass stripped from galaxies only contibutes a small fraction to the ICL (Murante et al. 2007). One prediction of the harassment model is that dwarf galaxies should lie on low surface brightness tidal streams or arcs. However, Davies et al (2007) have found no evidence for any such  features in a sample of dwarf galaxies in the Virgo cluster.

The most plausible explanation for the origin of the observed red dwarf population located in the innermost regions of nearby clusters is that they are the evolved blue dwarf galaxies observed in clusters at $z\approx 0.8$. Galaxy clusters are built through the continuous accretion of galaxies. These accreted galaxies can fall into the cluster potential either as a smooth rain of galaxies, or in small galaxy groups (McIntosh et al. 2004). These galaxy groups can contain a significant fraction of dwarf galaxies \citep[e.g., ][]{trentham06} and, given that ram-pressure stripping is not efficient in low-mass galaxy groups, these dwarfs can still contain a significant amount of gas and thus exhibit blue colors. Several strands of observational evidence indicate that galaxy clusters at high redshift show a larger fraction of substructure and mergers than  low redshift ones (van Dokkum et al. 1999; Halliday et al. 2004). Thus, the blue dwarf galaxies observed in clusters at high redshift could be recent cluster arrivals in small galaxy groups. These groups can then sink to the cluster center due to dynamical friction, leading to the presence of dwarf galaxies in the cluster innermost regions. Contrary to galaxy groups, ram-pressure stripping is quite important in massive clusters and dwarf galaxies can lose their gas content on small timescales ($\sim$ 50 Myr; Quilis et al. 2000). The loss of the gas reservoirs will produce an overall reddening of these galaxies, forming the red dwarf population that we observe in nearby clusters.

\acknowledgments

The authors would like to thank B. Moore, G. Yepes, and V. Quilis for useful discussion. We acknowledge  financial support by the Spanish Ministerio de Ciencia y Tecnolog\'{\i}a grants AYA2007-67965-C03-01


\begin{thebibliography}{}

\bibitem[Abell et al.(1989)]{abell89} Abell, G.O., Corwin, H.G., Jr.,\& Olowin, R.P.\ 1989, \apjs, 70, 1

\bibitem[Aguerri et al. 2007]{aguerri07} Aguerri, J.~A.~L.,S{\'a}nchez-Janssen, R. \& Mu{\~n}oz-Tu{\~n}{\'o}n, C., A\&A, 471, 17 (Paper I)

\bibitem[Aguerri et al. 2005a]{aguerri05a} Aguerri, J.~A.~L., 
Iglesias-P{\'a}ramo, J., V{\'{\i}}lchez, J.~M., Mu{\~n}oz-Tu{\~n}{\'o}n, 
C., \& S{\'a}nchez-Janssen, R.\ 2005a, \aj, 130, 475 

\bibitem[Aguerri et al. 2005b]{aguerri05b} Aguerri, J.~A.~L., 
Gerhard, O.~E., Arnaboldi, M., Napolitano, N.~R., Castro-Rodriguez, N., \& 
Freeman, K.~C.\ 2005b, \aj, 129, 2585 


\bibitem[Aguerri et al. 2004]{aguerri04} Aguerri, J.~A.~L., 
Iglesias-Paramo, J., Vilchez, J.~M., \& Mu{\~n}oz-Tu{\~n}{\'o}n, C.\ 2004, 
\aj, 127, 1344

\bibitem[Allam et al.(2005)]{allam05} Allam, S.~S., Tucker, 
D.~L., Lee, B.~C., \& Smith, J.~A.\ 2005, \aj, 129, 2062 

\bibitem[Arnaboldi et al. 2002]{arnaboldi02} Arnaboldi, M., et 
al.\ 2002, \aj, 123, 760 

\bibitem[Ashman et al.  1994]{ashman94} Ashman, K.~M., Bird, 
C.~M., \& Zepf, S.~E.\ 1994, \aj, 108, 2348

\bibitem[Binggeli et al. 1987]{binggeli87} Binggeli, B., Tammann, 
G.~A., \& Sandage, A.\ 1987, \aj, 94, 251 

\bibitem[Biviano \& Katgert  2004]{biviano04} Biviano, A., \& 
Katgert, P.\ 2004, \aap, 424, 779 

\bibitem[Biviano \& Girardi 2003]{biviano03} Biviano, A., \& 
Girardi, M.\ 2003, \apj, 585, 205 

\bibitem[Blanton et al. 2005]{blanton05} Blanton, M.~R., Lupton, 
R.~H., Schlegel, D.~J., Strauss, M.~A., Brinkmann, J., Fukugita, M., \& 
Loveday, J.\ 2005, \apj, 631, 208 

\bibitem[B{\"o}hringer et al.(2000)]{boringer00} B{\"o}hringer,
 H., et al.\ 2000, \apjs, 129, 435

\bibitem[Carlberg et al.  1997]{carlberg97} Carlberg, R.~G., Yee, 
H.~K.~C., \& Ellingson, E.\ 1997, \apj, 478, 462


\bibitem[Conselice 2002]{conselice02} Conselice, C.~J.\ 2002, 
\apjl, 573, L5 


\bibitem[Davies et al. 2007]{davies07} Davies, K. I., Roberts, S., \& Sabatini, S., 2007, MNRAS, in press

\bibitem[Dekel \& Silk 1986]{dekel86} Dekel, A., \& Silk, J.\ 
1986, \apj, 303, 39

\bibitem[De Lucia et al. 2007]{delucia07} De Lucia, G., et al.\ 
2007, \mnras, 374, 809 


\bibitem[de Rijcke et al. 2005]{derijcke05} de Rijcke, S., 
Michielsen, D., Dejonghe, H., Zeilinger, W.~W., \& Hau, G.~K.~T.\ 2005, 
\aap, 438, 491 

\bibitem[Goto  2005]{goto05} Goto, T.\ 2005, \mnras, 359, 1415

\bibitem[Halliday et al. 2004]{halliday04} Halliday, C., et al.\ 
2004, \aap, 427, 397 


\bibitem[Hogg et al. 2004]{hogg04} Hogg, D.~W., et al.\ 2004, 
\apjl, 601, L29 

\bibitem[Kodama et al. 2004]{kodama04} Kodama, T., et al.\ 
2004, \mnras, 350, 1005 

\bibitem[Lisker et al. 2006]{lisker06} Lisker, T., Grebel, 
E.~K., \& Binggeli, B.\ 2006, \aj, 132, 497 


\bibitem[Mastropietro et al. 2005]{mastropietro05} Mastropietro, C., 
Moore, B., Mayer, L., Debattista, V.~P., Piffaretti, R., \& Stadel, J.\ 
2005, \mnras, 364, 607 


\bibitem[McIntosh et al. 2004]{mcintosh04} McIntosh, D.~H., Rix, 
H.-W., \& Caldwell, N.\ 2004, \apj, 610, 161 

\bibitem[Merritt 1984]{merritt84} Merritt, D.\ 1984, \apj, 276, 
26 
 
\bibitem[Moore et al. 1996]{moore96} Moore, B., Katz, N., 
Lake, G., Dressler, A., \& Oemler, A.\ 1996, \nat, 379, 613 

\bibitem[Murante et al. 2007]{murante07} Murante, G., Giovalli, 
M., Gerhard, O., Arnaboldi, M., Borgani, S., \& Dolag, K.\ 2007, \mnras, 
377, 2 

\bibitem[Phillipps et al. 1998]{phillips98} Phillipps, S., 
Driver, S.~P., Couch, W.~J., \& Smith, R.~M.\ 1998, \apjl, 498, L119 

\bibitem[Popesso et al. 2006]{popesso06} Popesso, P., Biviano, 
A., B{\"o}hringer, H., \& Romaniello, M.\ 2006, \aap, 445, 29 

\bibitem[Quilis et al.  2000]{quilis00} Quilis, V., Moore, B., 
\& Bower, R.\ 2000, Science, 288, 1617 

\bibitem[Rakos \& Schombert 2004]{rakos04} Rakos, K., \& 
Schombert, J.\ 2004, \aj, 127, 1502

\bibitem[Stott et al.(2007)]{stott07} Stott, J.~P., Smail, I., 
Edge, A.~C., Ebeling, H., Smith, G.~P., Kneib, J.-P., \& Pimbblet, K.~A.\ 
2007, \apj, 661, 95 


\bibitem[Strateva et al.  2001]{strateva01} Strateva, I., et al.\ 
2001, \aj, 122, 1861

\bibitem[Trentham et al. 2006]{trentham06} Trentham, N., Tully, 
R.~B., \& Mahdavi, A.\ 2006, \mnras, 369, 1375 

\bibitem[Trujillo et al. 2002]{trujillo02} Trujillo, I., Aguerri, 
J.~A.~L., Guti{\'e}rrez, C.~M., Caon, N., \& Cepa, J.\ 2002, \apjl, 573, L9 

\bibitem[White \& Rees 1978]{white78} White, S.~D.~M., \& 
Rees, M.~J.\ 1978, \mnras, 183, 341 

\bibitem[van Dokkum et al. 1999]{vandokkum99} van Dokkum, P.~G., 
Franx, M., Fabricant, D., Kelson, D.~D., \& Illingworth, G.~D.\ 1999, 
\apjl, 520, L95 


\bibitem[Voges et al.(1999)]{voges99} Voges, W., et al.\ 1999,
 \aap, 349, 389
 

\bibitem[York et al. 2000]{york00} York, D.~G., et al.\ 2000, \aj, 120, 1579

\bibitem[Zabludoff et al. 1990]{zabludoff90} Zabludoff, A.~I., 
Huchra, J.~P., \& Geller, M.~J.\ 1990, \apjs, 74, 1

\bibitem[Zwicky et al.(1961)]{zwicky61} Zwicky, F., Herzog, E.,
 \& Wild, P.\ 1961, Pasadena: California Institute of Technology (CIT),
 |c1961,
\end{thebibliography}
\end{document}